# Core–Shell Structured Dielectric–Metal Circular Nanodisk Antenna: Gap Plasmon Assisted Magnetic Toroid-like Cavity Modes

*Qiang Zhang,[†] Jun Jun Xiao,[†,*] Xiao Ming Zhang,[†] Dezhuan Han,[‡] and Lei Gao[§]*

[†]College of Electronic and Information Engineering, Shenzhen Graduate School, Harbin Institute of Technology, Shenzhen 518055, China

[‡]Department of Applied Physics, Chongqing University, Chongqing 400044, China

[§]Jiangsu Key Laboratory of Thin Films, School of Physical Science and Technology, Soochow University, Suzhou 215006, China

**ABSTRACT:** Plasmonic nanoantennas, the properties of which are essentially determined by their resonance modes, are of interest both fundamentally and for various applications. Antennas with various shapes, geometries and compositions have been demonstrated, each possessing unique properties and potential applications. Here, we propose the use of a sidewall coating as an additional degree of freedom to manipulate plasmonic gap cavity modes in strongly coupled metallic nanodisks. It is demonstrated that for a dielectric middle layer with a thickness of a few tens of nanometers and a sidewall plasmonic coating of more than ten nanometers, the usual optical magnetic resonance modes are eliminated, and only magnetic toroid-like modes are sustainable in the infrared and visible regime. All of these deep-subwavelength modes can be interpreted as an interference effect from the gap surface plasmon polaritons. Our results will be useful in nanoantenna design, high-Q cavity sensing, structured light-beam generation, and photon emission engineering.

Surface plasmon polaritons (SPPs) are known as the collective oscillations of conductive electrons at metallic nanostructures and their interfaces.[1,2] When the optical field couples with the SPPs in plasmonic nanostructures, some fascinating features and applications arise, such as strong local field enhancement,[3-5] imaging beyond the diffraction limit,[6-10] and extraordinary transmission through periodic arrays of subwavelength holes in optically thick metallic films.[11,12] Indeed, any plasmonic nanostructure can be regarded as an optical nanoantenna or plasmonic resonator because of its ability to radiate and collect light, similar to the behavior of RF antennas in the microwave regime.[13-16] Moreover, because of the deep-subwavelength localization and strong enhancement of the local fields, plasmonic antennas have attracted considerable attention for their promising potential applications in areas such as optoelectronic integrated circuits,[17,18] surface-enhanced Raman scattering detection,[19-22] plasmonic sensing,[23-25] optical micro-manipulation,[26,27] and light harvesting.[28-30]

Recently, circular metal–dielectric–metal resonators (CMDMRs) have been studied as plasmonic microcavities for various applications because of their extremely high Q factor and low modal volume.[31-38] The field patterns of the resonance modes in a CMDMR are similar to the whispering gallery modes (WGM) in a high-refractive-index dielectric disk. Therefore, these modes are known as WGM-like plasmonic cavity modes in the literature.[34,35] More notably, the in-plane magnetic field of one of the basic resonance modes in a CMDMR possesses a vortex distribution and has been interpreted as a predominant magnetic toroidal dipole.[37,38] This behavior is essentially enabled by the plasmon-resonance-induced displacement currents, which can overcome the saturation of the conduction current and promote the optical magnetic resonance from the terahertz up to the visible regime.[39] Filter *et al.* have established an analytical theory to explore the resonance properties of CMDMRs by defining a reflection coefficient of the gap SPPs at the termination of such a circular antenna.[40] It has been demonstrated that the optical field in the gap region of a CMDMR originates from the superimposed cylindrical gap SPPs. This type of coherent gap SPP is termed a Hankel-type or Bessel-type plasmonic standing wave.[40,41] The resonance modes in a CMDMR can be determined by solving the scalar Helmholtz equation in the structure with appropriate boundary conditions. At the lateral interface between the interior dielectric layer and the exterior air, the boundary condition is essentially the Neumann condition, e.g., $\partial E_z/\partial \rho = 0$, where $\rho$ is the radial component in cylindrical coordinates.[36]

In this work, we demonstrate that adding a sidewall metallic coating to a CMDMR can strongly affect the mode pattern and the optical characteristics of the cavity modes. Intuitively, the sidewall coating transforms the CMDMR from a sandwiched structure into a core–shell structured nanodisk (CSND), representing a dramatic change (from an open cavity to a closed one) in geometrical topology (see Figure 1). The cavity modes can be easily excited by an incident wave in a configuration that allows the magnetic component to cross the gap horizontally (see Section 1 of the Supporting Information for details). However, unlike the open cavity of a CMDMR, all cavity modes in a CSND have magnetic fields with vortex patterns. That is, the in-plane magnetic fields are purely combinations of magnetic vortices, making them similar to the hybridized modes of magnetic toroidal dipoles. We further demonstrate that all these toroid-like cavity modes follow the dispersion relation of the gap SPPs. The resonance condition is approximately given by the Dirichlet boundary condition $E_z(\rho) = 0$ at the edge of the dielectric layer.

Figure 1 schematically illustrates the CSND, which consists of a concentric $SiO_2$ circular disk and a silver coating. The permittivity of $SiO_2$ is set to $\varepsilon_d = 2.1$, and that of silver, $\varepsilon_m$, is taken from Johnson and Christy.[42] To examine the cavity modes in the CSND, we consider a $z$-polarized plane wave propagating along the $x$ direction that is incident upon the CSND. We employed two different methods — the finite integration technique (FIT) and the finite element method (FEM) — to calculate the optical cross sections.[43,44] The results are presented in Figure 2a for $R = 260$ nm, $d = 20$ nm, $D_z = 35$ nm, and $D_x = 40$ nm. In the frequency band of interest, seven significant peaks are observed in the absorption cross section (ACS; see the blue dash-dotted line). The frequencies of these peaks are $f = 156$ THz ($\lambda = 1923$ nm), 245 THz (1224 nm), 322 THz (932 nm), 344 THz (872 nm), 391 THz (767 nm), 423 THz (709 nm), and 489 THz (613 nm). These peaks are also observable in the extinction cross section (ECS) and the scattering cross section (SCS). For the other two wave-excitation configurations, the plasmonic electric multipole modes dominate the optical response, and these gap cavity modes are so nearly "dark" as to be unobservable (see Figures S1 and S2 in the Supporting Information).

To identify these cavity modes, we plot the $z$ component of the electric field, $E_z$, and the in-plane magnetic field vector in the middle slice ($z = 0$ nm) of the structure in Figures 3a and 3b, respectively. Figure 3b reveals that the magnetic fields of all these cavity modes have vortex distributions. For the first mode at $f = 156$ THz, the magnetic field is confined in a circular form. This mode is, in fact, a typical magnetic toroidal dipole mode with a dominant toroidal moment in the $z$ direction.[37,38] For the remaining higher-order cavity modes, the magnetic field can consist of several vortices arranged in different fashions. Therefore, we regard these resonance cavity modes as toroid-like modes and use ad-hoc terminology, labeling them **T** modes to differentiate them from conventional magnetic modes. One can distinguish these modes by their azimuthal and radial numbers in the $E_z$ field patterns (see Figure 3a). For example, the azimuthal and radial numbers of the first mode are "0" and "1," respectively. Thus, this mode can be labeled **T**(0,1). All seven cavity modes in Figure 2a are labeled in a similar manner.

To elucidate the effects of the sidewall plasmonic coating on the field distributions of the cavity modes, the results of the no-coating situation (open-cavity case), i.e., the results for a CMDMR with the same geometry and $D_x = 0$, are also presented. The calculated optical spectra are presented in Figure 2b. Seven remarkable peaks are also observed in this case, at $f = 124$ THz (2419 nm), 204 THz (1470 nm), 248 THz (1209 nm), 277 THz (1083 nm), 338 THz (887 nm), 429 THz (699 nm), and 499 THz (601 nm). The $E_z$ and in-plane magnetic field patterns are presented in Figures 3c and 3d, respectively. Comparing Figures 3a and 3c, we observe that some of the cavity modes have nearly identical $E_z$ distributions. For example, the pattern for $f = 124$ THz in Figure 3c and that for $f = 245$ THz in Figure 3a have the same azimuthal and radial numbers (1 and 1, respectively). However, the corresponding magnetic fields are quite different. The magnetic field at $f = 124$ THz in Figure 3d and that at $f = 245$ THz in Figure 3b are distinct: the latter exhibits a double-circle pattern, whereas the former exhibits a typical magnetic dipole distribution. Based on the magnetic field distributions depicted in Figure 3d, it is convenient to divide the cavity modes in the CMDMR into two categories. If the magnetic field of a mode does not form any vortices within the cavity, we can treat the field as a magnetic multipole (labeled with **M**). We regard the remaining modes as **T** modes, which also appear in the CSND case. Based on both the azimuthal and radial numbers represented in Figure 3c, all seven modes in the CMDMR can be labeled as **M**(m,n) or **T**(m,n), as demonstrated in Figure 2b. It is important to note that none of the **M** modes survives when the sidewall plasmonic coating is present. This finding indicates that one of the effects of the sidewall coating is to eliminate the magnetic multipole cavity modes. Essentially, the **T**-mode and **M**-mode currents are different.

The sidewall plasmonic coating prevents the conduction current from terminating on the upper or lower disk edge. Such edge termination of the current is crucial to the formation of **M** modes (see Figure S3 in the Supporting Information). We replaced the silver of the sidewall coating with PEC, and the results were nearly the same except for a small shift in the resonance frequencies (see Figure S4 in the Supporting Information). This finding strongly suggests that the sidewall provides a crucial bridge between the upper and lower disks for the induced current, which prevents the formation of the usual **M** modes in the CSND.

Another remarkable characteristic of the ECS (black solid lines) and SCS (red dashed lines) spectra shown in Figures 2a and 2b is that the resonance peaks of these cavity modes lie on the shoulder of a much broader resonance peak. We demonstrate that this broad resonance originates from the SPP resonance of the entire metallic shell by examining the near-field distribution and the current density at this frequency (see Figures S6 and S7 in the Supporting Information). The resonance occurs at much higher frequency (~740 THz), which is outside the range of Figure 2. This SPP exists at the interface between the metallic shell and the background (air). For convenience, we call this SPP an "exterior SPP" to distinguish it from the gap SPPs sustained in the dielectric core layer. Note that the upper and lower disk thicknesses are sufficiently large to disallow strong and direct evanescent wave coupling between the exterior SPP and the gap SPPs. However, the scattering interference between the exterior SPP mode and the relatively sharp gap SPP modes leads to an asymmetric spectrum, featuring a Fano profile (see Figure 2).

Based on the above results, we infer that the apparent differences in mode characteristics between the CSND and the CMDMR must result from the difference in the boundary condition in terms of the gap SPPs. To clarify this point, we examine the electric field at the edges of the dielectric layer. Two corresponding pairs of modes in the CSND and CMDMR are selected as examples: one corresponding to **T**(1,1) in the CSND and **M**(1,1) in the CMDMR and the other corresponding to **T**(0,1) in the CSND and **T**(0,2) in the CMDMR. The two members of the former pair have the same azimuthal and radial numbers (1 and 1), whereas those of the latter pair have similar single-vortex magnetic fields (as observed in Figures 3b and 3c). For all four modes, the magnitude of $E_z(x)$ along the $x$ coordinate is shown in Figure 4. The two vertical dotted lines indicate the boundaries ($\rho = R$) of the dielectric layer at $x_0 = \pm 260$ nm. It is clearly observed that for the CMDMR, the field $E_z(x = x_0)$ is always close to its local maximum (dashed curves), whereas for the CSND, $E_z(x = x_0)$ tends to be vanishing (solid curves). Moreover, if we use PEC as the material for the sidewall coating, then $E_z(x = x_0)$ must be strictly equal to zero (see Figure S5 in the Supporting Information). For the non-coated case (the open cavity), the Neumann boundary condition $\partial E_z(R)/\partial \rho = 0$ can be applied as an approximation, as reported in Ref. [36]. In the meantime, we expect the Dirichlet boundary condition, $E_z(R) = 0$, to be appropriate in the coated case (the closed cavity).

In a circular patch nanoantenna, the solution to the Helmholtz equation for the $z$ component of the electric field can be written in cylindrical coordinates:[36]

$$E_z(\rho, \phi, z) = a(z)[H_m^{(1)}(k_{gsp}\rho) + r_m H_m^{(2)}(k_{gsp}\rho)]e^{im\phi} \quad (1)$$

where $m$ is an integer (angular quantum number), $H_m^{(1,2)}$ is the $m$-th Hankel function of the first or second type, $k_{gsp}$ is the wave vector of the gap SPPs in the metal–dielectric–metal sandwich structure, and $r_m$ represents the complex reflection coefficient of the Hankel-type SPP from the sidewall (lateral interface). In Eq. (1), $a(z)$ is the evanescent modal profile in the $z$ direction and $\varphi$ is the rotational angle. For the CMDMR, the Neumann boundary condition $\partial \vec{E}/\partial \rho = 0$ is confirmed by plotting the resonance frequencies of the cavity modes versus $\chi'_{mn}/(R+d)$, where $\chi'_{mn}$ represents the n-th zero point of the derivative of the m-th Bessel function $J'_{mn}(k_{gsp}\rho)$.[36] In the CSND, the lateral boundary becomes a dielectric–metal interface, and the electric field nearly vanishes at $\rho = R$ (see Figure 4). It is expected that the Dirichlet boundary condition $E_z(\rho = R) = 0$ should approximately apply. We plotted the resonance frequencies of the cavity modes in the CSND versus $\chi_{mn}/R_{eff}$; here, $\chi_{mn}$ is the $n$-th zero point of the $m$-th Bessel function $J_{mn}(k_{gsp}\rho)$. The effective radius can be approximated to first order as $R_{eff} = R + d$. As observed in Figure 5, the resonance points (symbols) of the considered seven toroid-like modes all fall on the curve of the dispersion relation (thick solid curve) of the gap SPPs, $\omega(k_{gsp})$. For simplicity, we assume that $D_z \to \infty$ and $k_{gsp}d \ll 1$, yielding the following approximate expression for the dispersion relation of the gap SPPs:[45,46]

$$k_{gsp} \approx k_0\sqrt{\varepsilon_d + 0.5(k_{gsp}^0/k_0)^2 + \sqrt{(k_{gsp}^0/k_0)^2\left[\varepsilon_d - \varepsilon_m + 0.25(k_{gsp}^0/k_0)^2\right]}} \qquad (2)$$

In Eq. (2), $k_0$ is the vacuum wave vector and $k_{gsp}^0 = -2\varepsilon_d/d\varepsilon_m$ is the wave vector of the gap SPPs in the limit of vanishing gap thickness ($d \to 0$). The results presented in Figure 5 confirm that the gap surface plasmons with the lateral Dirichlet boundary condition are responsible for the emergence of the toroidal and toroid-like cavity modes. The ACS versus the disk size $R$ in the frequency band from 100 THz to 500 THz, which quantifies the relative strength of the excitation, is plotted in Figure S8 in the Supporting Information.

Finally, the dependences of the resonance frequency on the geometric parameters $d$, $D_x$ and $D_z$ are in order. Figure 6a demonstrates that the resonance frequencies of all cavity toroid-like modes blueshift as the gap thickness $d$ increases. In fact, this blueshift effect is also captured by the dispersion relation of the gap SPPs.[46] The dispersion curve governed by Eq. (2) and depicted in Figure 5 would ascend for increasing $d$. In an equivalent *LC* circuit model, a thicker dielectric layer results in smaller capacitance and causes the magnetic-field-induced anti-parallel currents over the upper and lower disks to oscillate at a higher frequency. For the thickness $D_x$ of the sidewall plasmonic coating, Figure 6b shows that the resonance frequencies remain nearly unchanged except in the case of a very thin metallic layer (e.g., $D_x < 20$ nm). This finding is observed because the electromagnetic field leaks laterally toward the exterior and interacts with the background field when the sidewall coating is not sufficiently thick. This field leaky effect could be utilized in sensing applications because the cavity modes can be dramatically affected by outside perturbations near the CSND surface. The thickness $D_z$ of the metallic layer in the $z$ direction plays a key role in determining the effective refractive index of the gap SPPs, the dispersion relation, and the mode profile. Figure 6c demonstrates that the resonance frequencies are insensitive to $D_z$ variations as long as $D_z > 30$ nm, justifying the use of Eq. (2) in such cases. However, for $D_z < 30$ nm, the dispersion relation represented by Eq. (2) deviates from the numerical data and must be replaced with a five-layer result in which the thickness of the metallic layer is finite and is appropriately considered.[46] Nevertheless, for larger $D_z$, the excitation efficiency of some cavity modes becomes too weak to observe (see Figure S9 in the Supporting Information). The small deviations observed in the high-frequency (high k-vector) regime of Figure 5 are partially attributable to the inaccuracy of Eq. (2) at small disk thicknesses, e.g., $D_z = 35$ nm in our case.

In a real application such as photoluminescence (PL) enhancement or spontaneous-emission control, a substrate must necessarily be present, and an active material covering (e.g., MEH-PPV) may be employed (see Figure S10 in the Supporting Information). In such a case, we demonstrate that it is possible to shift the exterior SPP frequency further away from those of the gap SPP modes because the substrate can substantially affect the exterior SPP. In the meantime, the gap SPPs can remain intact. In particular, by virtue of field symmetry, it is demonstrated that for a dipole emitter located on the center axis of the cavity, the excitation of toroid modes with an azimuthal number of 0 becomes much stronger compared with the other modes (see Figures S11 and S12 in the Supporting Information). This effect provides a feasible approach to the exploration of the coupling dynamics between a quantum emitter and the magnetic toroidal moment by measuring the spectral and temporal PL spectra. The strain and conformation effects in the luminescent slab could also be interesting topics and require further exploration.

In conclusion, we theoretically and numerically investigated the resonance properties of the cavity modes in core–shell structured dielectric–metal nanodisk antennas. The magnetic fields of all cavity modes in such resonators are observed to be dominated by vortex patterns, in either ring form or side-by-side form. Compared with the case of a conventional metal–dielectric–metal circular nanodisk antenna, we revealed that magnetic multipole cavity modes are eliminated by the core–shell structure. The plasmonic sidewall coating forces the electric fields at the lateral edge of the dielectric layer to nearly vanish and provides a bridge connecting the conduction currents in the upper and lower disks. All these deep-subwavelength cavity modes can be interpreted as a strong interference effect from the gap SPPs. The unique optical field distributions and the resonance properties in the CSND may be useful in exploiting toroidal metamaterials. Our results provide a new platform to explore the interactions of both fundamental and high-order toroidal modes and may find application in nanoantenna design, high-Q cavity sensing, structured light-beam generation, and photon emission engineering.

## ASSOCIATED CONTENT

**Supporting Information**

Responses for incident waves in the other two configurations; Current density distributions in CSND and CMDMR; CSND with perfect dielectric conductor (PEC) sidewall coating; Fano effects from the interference between the exterior SPP and the gap SPP; Absorption spectra for various dielectric disk radii; Absorption spectra for various thicknesses of the metal layers; Decay rates of an electric point dipole near the open and closed cavities. This material is available free of charge via the Internet at http://pubs.acs.org.

## AUTHOR INFORMATION

**Corresponding Author**

*E-mail: eiexiao@hitsz.edu.cn.

**Notes**

The authors declare no competing financial interest.

## ACKNOWLEDGMENTS

Stimulating discussions with C. T. Chan and X. Y. Deng are greatly appreciated. This work was supported by the NSFC (Grant Nos. 11274083, 11004043, 11304038, and 11374223), the Shenzhen Municipal Science and Technology Plan (Grant Nos. KQCX20120801093710373, JCYJ20120613114137248, and 2011PTZZ048), the National Basic Research Program under Grant No. 2012CB921501, and the project funded by the Priority Academic Program Development of Jiangsu Higher Education Institutions. We acknowledge assistance from the National Supercomputing Center in Shenzhen.

## REFERENCES

(1) Raether, H. *Surfacep plasmons on smooth and rough surfaces and on gratings*; Springer, New York, 1988.
(2) Pitarke, J. M.; Silkin, V. M.; Chulkov, E. V.; Echenique, P. M. Theory of surface

plasmons and surface-plasmon polaritons. *Rep. Prog. Phys.* **2007**, *70*, 1–87.

(3) Boyd, G. T.; Rasing, Th.; Leite, J. R. R.; Shen, Y. R. Local-field enhancement on rough surfaces of metals, semimetals, and semiconductors with the use of optical second-harmonic generation. *Phys. Rev. B* **1984**, *30*, 519-526.

(4) Klar, T.; Perner, M.; Grosse, S.; von Plessen, G.; Spirkl, W.; Feldmann, J. Surface-plasmon resonances in single metallic nanoparticles. *Phys. Rev. Lett.* **1998**, *80*, 4249-4252.

(5) Lu, Y.; Liu, G. L.; Kim, J.; Mejia, Y. X.; Lee, L. P. Nanophotonic crescent moon structures with sharp edge for ultrasensitive biomolecular detection by local electromagnetic field enhancement effect. *Nano Lett.* **2005**, *5*, 119-124.

(6) Gramotnev, D. K.; Bozhevolnyi, S. I. Plasmonics beyond the diffraction limit. *Nat. Photon.* **2010**, *4*, 83-91.

(7) Barnes, W. L. Surface plasmon–polariton length scales: a route to sub-wavelength optics. *J. Opt. A: Pure Appl. Opt.* **2006**, *8*, S87–S93.

(8) Maier, S. A.; Kik, P. G.; Atwater, H. A.; Meltzer, S.; Harel, E.; Koel, B. E.; Requicha, A. A. G. Local detection of electromagnetic energy transport below the diffraction limit in metal nanoparticle plasmon waveguides. *Nat. Mater.* **2003**, *2*, 229-232.

(9) Barnes, W. L.; Dereux, A.; Ebbesen, T. W. Surface plasmon subwavelength optics. *Nature* **2003**, *424*, 824-830.

(10) Schuller, J. A.; Barnard, E. S.; Cai, W.; Jun, Y. C.; White, J. S.; Brongersma, M. L. Plasmonics for extreme light concentration and manipulation. *Nat. Mater.* **2010**, *9*, 193-204.

(11) Barnes, W. L.; Murray, W. A.; Dintinger, J.; Devaux, E.; Ebbesen, T. W. Surface plasmon polaritons and their role in the enhanced transmission of light through periodic arrays of subwavelength holes in a metal film. *Phys. Rev. Lett.* **2004**, *92*, 107401.

(12) Ebbesen, T. W.; Lezec, H. J.; Ghaemi, H. F.; Thio, T.; Wolff. P. A. Extraordinary optical transmission through sub-wavelength hole arrays. *Nature* **1998**, *391*, 667-669.

(13) Alù, A.; Engheta, N. Hertzian plasmonic nanodimer as an efficient optical nanoantenna. *Phys. Rev. B* **2008**, *78*, 195111.

(14) Bharadwaj, P.; Deutsch, B.; Novotny, L. Optical antennas. *Adv. Opt. Photon.* **2009**, *1*, 438-483.

(15) Alù, A.; Engheta, N. Wireless at the nanoscale: optical interconnects using matched

nanoantennas. *Phys. Rev. Lett.* **2010**, *104*, 213902 (1-4).

(16) Esteban, R.; Teperik, T. V.; Greffet, J. J. Optical patch antennas for single photon emission using surface plasmon resonances. *Phys. Rev. Lett.* **2010**, *104*, 026802.

(17) Large, N.; Abb, M.; Aizpurua, J.; Muskens, O. L. Photoconductively loaded plasmonic nanoantenna as building block for ultracompact optical switches. *Nano Lett.* **2010**, *10*, 1741-1746.

(18) Huang, J. S.; Feichtner, T.; Biagioni, P.; Hecht, B. Impedance matching and emission properties of nanoantennas in an optical nanocircuit. *Nano Lett*. **2009**, *9*, 1897-1902.

(19) Kneipp, K.; Wang, Y.; Kneipp, H.; Perelman, L. T.; Itzkan, I. Single molecule detection using surface-enhanced Raman scattering (SERS). *Phys. Rev. Lett.* **1997**, *78*, 1667-1670.

(20) Smythe, E. J.; Dickey, M. D.; Bao, J.; Whitesides, G. M.; Capasso, F. Optical antenna arrays on a fiber facet for in situ surface-enhanced Raman Scattering detection. *Nano Lett*. **2009**, *9*, 1132-1138.

(21) Jackson, J. B.; Westcott, S. L.; Hirsch, L. R.; West, J. L.; Halas, N. J. Controlling the surface enhanced Raman effect via the nanoshell geometry. *App. Phys. Lett*. **2003**, 82, 257-259.

(22) Nie, S.; Emory, S. R.; Probing single molecules and single nanoparticles by surface-enhanced Raman Scattering. *Science* **1997**, *275*, 1102-1106.

(23) Lal, S.; Link, S.; Halas, N. J. Nano-optics from sensing to waveguiding. *Nat. Photon.* **2007**, *1*, 641-648.

(24) Liu, N.; Mesch, M.; Weiss, T.; H*entschel, M.*; **Gie**ssen, H. Infrared perfect absorber and its application as plasmonic sensor. *Nano Lett.* **2010**, *10*, 2342-2348.

(25) Bozhevolnyi, S. I.; Volkov, V. S.; Devaux, E.; Laluet, J. Y.; Ebbesen, T. W. Channel plasmon subwavelength waveguide components including interferometers and ring resonators. *Nature* **2006**, *440*, 508-511.

(26) Juan, M. L.; Righini, M.; Quidant, R. Plasmon nano-optical tweezers. *Nat. Photon.* **2011**, *5*, 349-356.

(27) Zhang, Q.; Xiao, J. J.; Zhang, X. M.; Yao, Y.; Liu, H. Reversal of optical binding force by Fano resonance in plasmonic nanorod heterodimer. *Opt. Express* **2013**, *5*, 6601-6608.

(28) Catchpole, K. R.; Polman, A. Plasmonic solar cells. *Opt. Express* **2008**, *16*, 21793-21800.

(29) Hägglund, C.; Apell, S. P. Plasmonic near-field absorbers for ultrathin solar cells. *J. Phys.*

*Chem. Lett.* **2012**, *3*, 1275-1285.

(30) Cui, Y.; Xu, J.; Fung, K. H.; Jin, Y.; Kumar, A.; He, S.; Fang, N. X. A thin film broadband absorber based on multi-sized nanoantennas. *App. Phys. Lett.* **2011**, *99*, 253101(1-4).

(31) Dmitriev, A.; Pakizeh, T.; Käll, M.; Sutherland, D. S. Gold–silica–gold nanosandwiches: tunable bimodal plasmonic resonators. *Small* **2007**, *3*, 294-299.

(32) Frederiksen, M.; Bochenkov, V. E.; Ogaki, R.; Sutherland, D. S. Onset of bonding plasmon hybridization preceded by gap modes in dielectric splitting of metal disks. *Nano Lett.* **2013**, *13*, 6033-6039.

(33) Kuttge, M.; F.; de Abajo, J. G.; Polman, A. Ultrasmall mode volume plasmonic nanodisk resonators. *Nano Lett*. **2010**, *10*, 1537-1541.

(34) Kwon, S. H. Deep subwavelength plasmonic whispering gallery-mode cavity. *Opt. Express* **2012**, *20*, 24918-24924.

(35) Lou, F.; Yan, M.; Thylen, L.; Qiu, M.; Wosinski, L. Whispering gallery mode nanodisk resonator based on layered metal-dielectric waveguide. *Opt. Express* **2014**, *22*, 8490-8502.

(36) Minkowski, F.; Wang, F.; Chakrabarty, A.; Wei, Q. H. Resonant cavity modes of circular plasmonic patch nanoantennas. *App. Phys. Lett.* **2014**, *104*, 021111.

(37) Dong, Z. G.; Zhu, J.; Yin, X.; Li, J.; Lu, C.; Zhang, X. All-optical Hall effect by the dynamic toroidal moment in a cavity-based metamaterial. *Phys. Rev. B* **2013**, *87*, 245429(1-5).

(38) Zhang, Q.; Xiao, J. J.; Wang, S. L. Optical characteristics associated with magnetic resonance in toroidal metamaterials of vertically coupled plasmonic nanodisks. *J. Opt. Soc. Am. B* **2014**, *31*, 1103-1108.

(39) Nazir, A.; Panaro, S.; Zaccaria, R. P.; Liberale, C.; Angelis F. D.; Toma, A. Fano coil-type resonance for magnetic hot-spot generation. *Nano Lett.* **2014**, *14*, 3166-3171.

(40) Filter, R.; Qi, J.; Rockstuhl, C.; Lederer, F. Circular optical nanoantennas: an analytical theory. *Phys. Rev. B* **2012**, *85*, 125429.

(41) Nerkararyan, S.; Nerkararyan, K. Generation of Hankel-type surface plasmon polaritons in the vicinity of a metallic nanohole. *Phys. Rev. B* **2010**, *82*, 245405.

(42) Johnson, P. B.; Christy, R. W. The optical constants of noble metals. *Phys. Rev. B* **1972**, *6*, 4370-4379.

(43) http://www. cst.com.

(44) http://www.comsol.com.

(45) Bozhevolnyi, S. I.; Jung, J. Scaling for gap plasmon based waveguides. *Opt. Express* **2008**, *16*, 2676-2684.

(46) Economou, E. N. Surface plasmons in thin films. *Phys. Rev.* **1969**, *182*, 539-554.

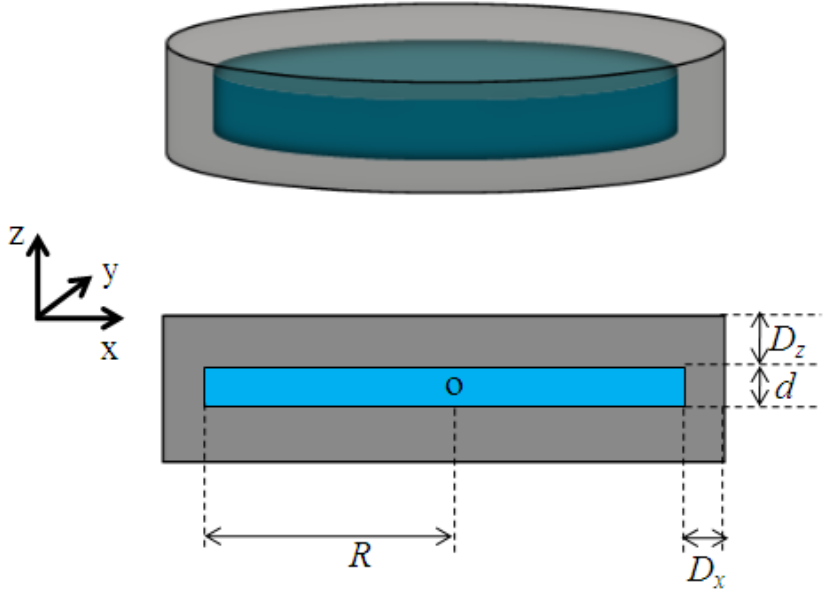

**Figure 1.** Schematic illustration of the geometry of the core–shell-structured nanodisk (CSND). An SiO$_2$ nanodisk of radius $R$ and thickness $d$ is coated with silver layers. The thickness of the metallic layer in the $z$ direction is $D_z$, and that in the $x$ direction is $D_x$. The origin of the coordinate system is placed at the center of the SiO$_2$ disk.

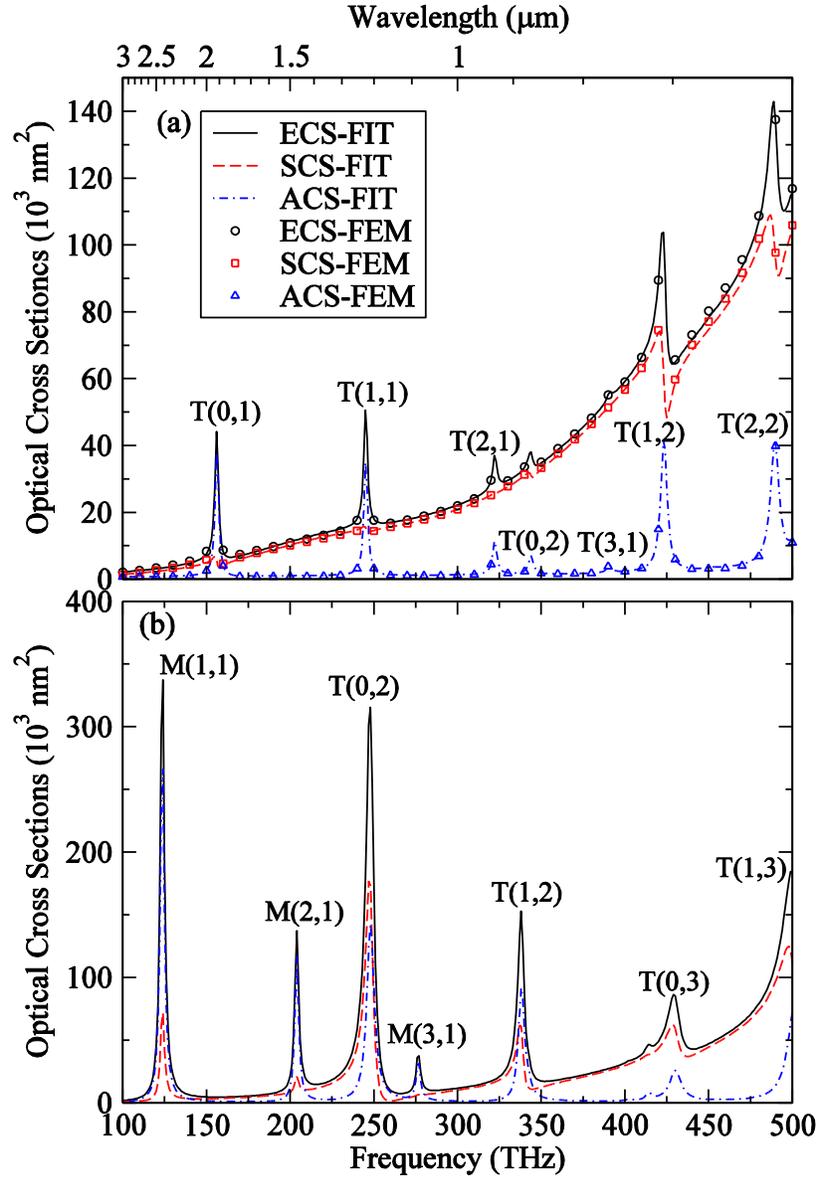

**Figure 2.** (a) Optical cross sections of the CSND with $R = 260$ nm, $d = 20$ nm, $D_x = 40$ nm and $D_z = 35$ nm calculated using two numerical methods (the lines represent the CST Microwave Studio results, and the symbols represent the COMSOL Multiphysics results). The **T** modes are labeled at the corresponding peaks of the spectrum. (b) Optical cross sections of a CMDMR with the same geometry for $D_x = 0$ nm. The cavity modes are categorized into **M** and **T** modes based on their in-plane magnetic field characteristics.

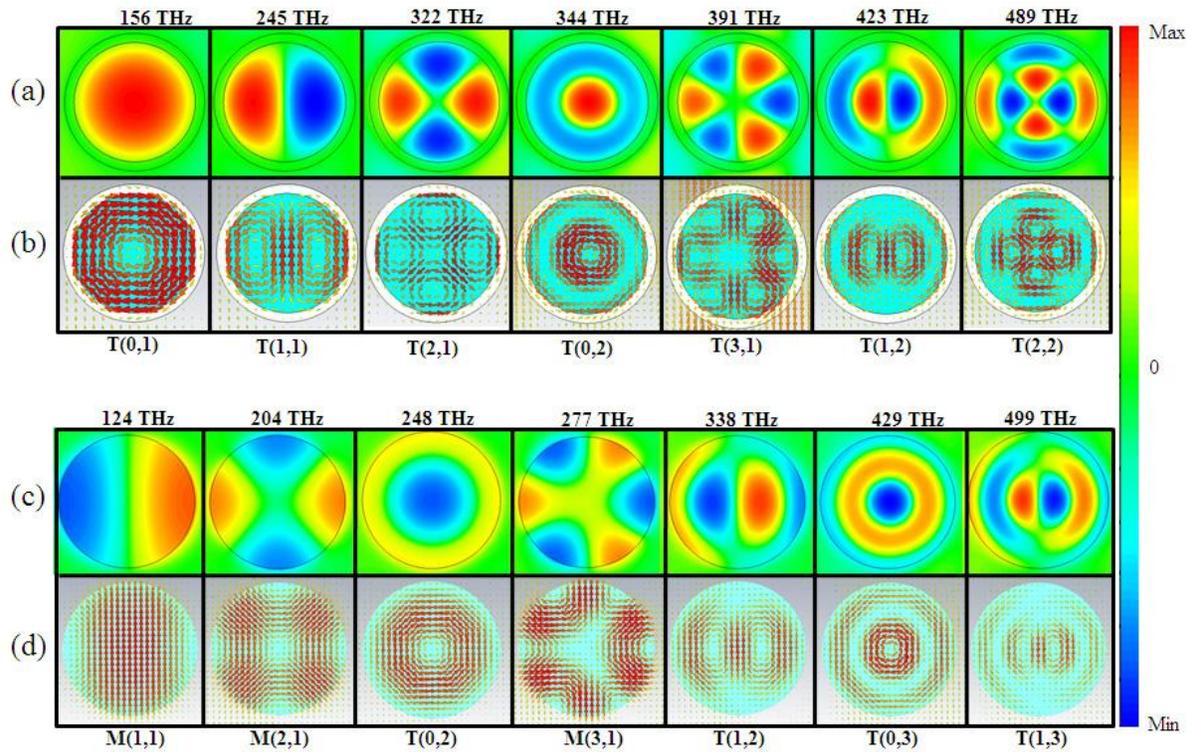

**Figure 3.** (a) Patterns of the electric fields $E_z$ of the seven cavity modes of the CSND. (b) The magnetic field distributions of the corresponding modes of the CSND. (c) Field patterns $E_z$ of the seven cavity modes of the CMDMR. (d) The magnetic field distributions of the corresponding modes of the CMDMR. The resonance frequency of each mode is indicated above the patterns, and their labels are provided below the patterns. The color bar is normalized to the full data range in each case.

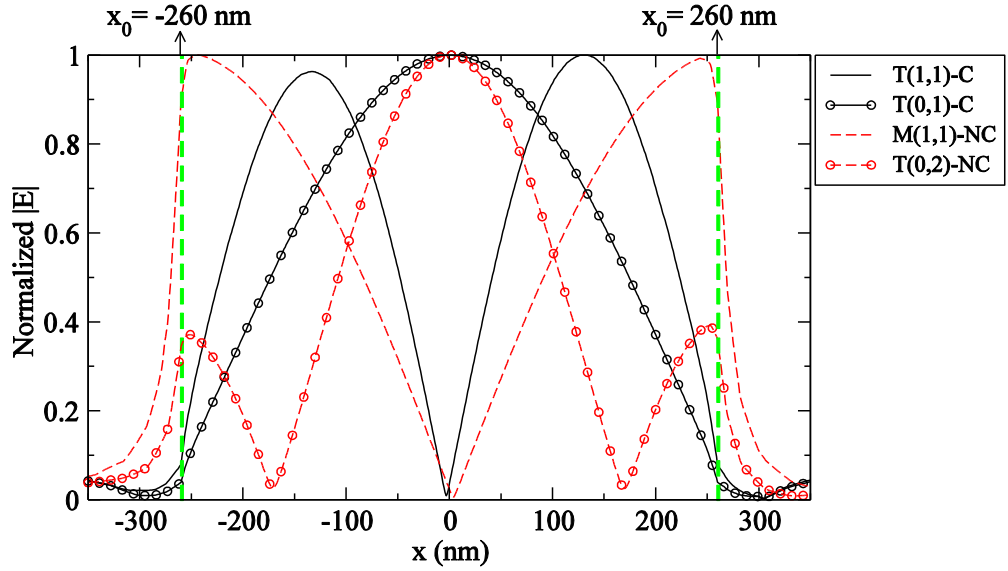

**Figure 4.** The electric field along the radial direction, normalized to the corresponding maximum. The vertical green dashed lines indicate the edges of the SiO$_2$ at $x_0 = \pm 260$ nm. The intersection points of the green dashed lines with the red dashed lines (for the CMDMR) are close to the local maxima, whereas those with the black solid lines (for the CSND) are close to zero.

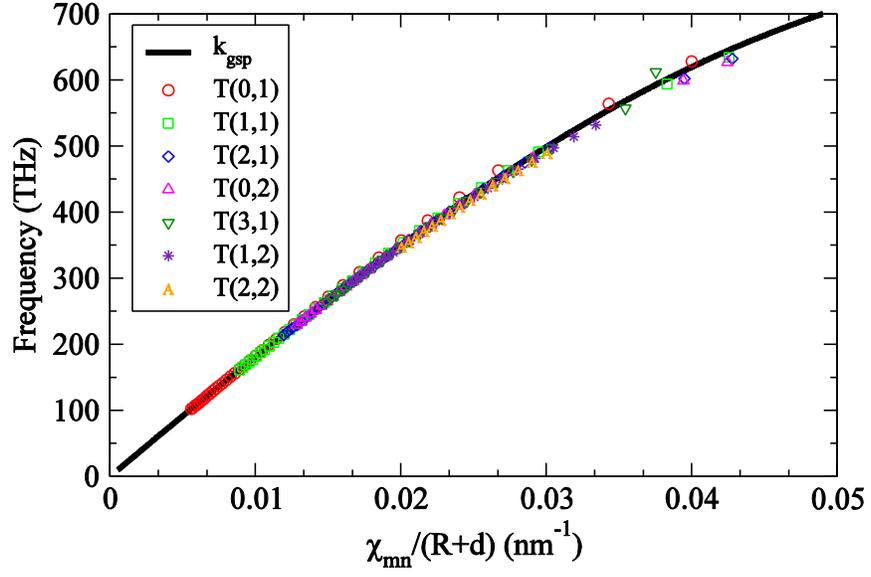

**Figure 5.** The resonance frequencies versus $\chi_{mn}/(R+d)$. The black solid line represents the dispersion relation of the gap SPPs obtained from Eq. (2).

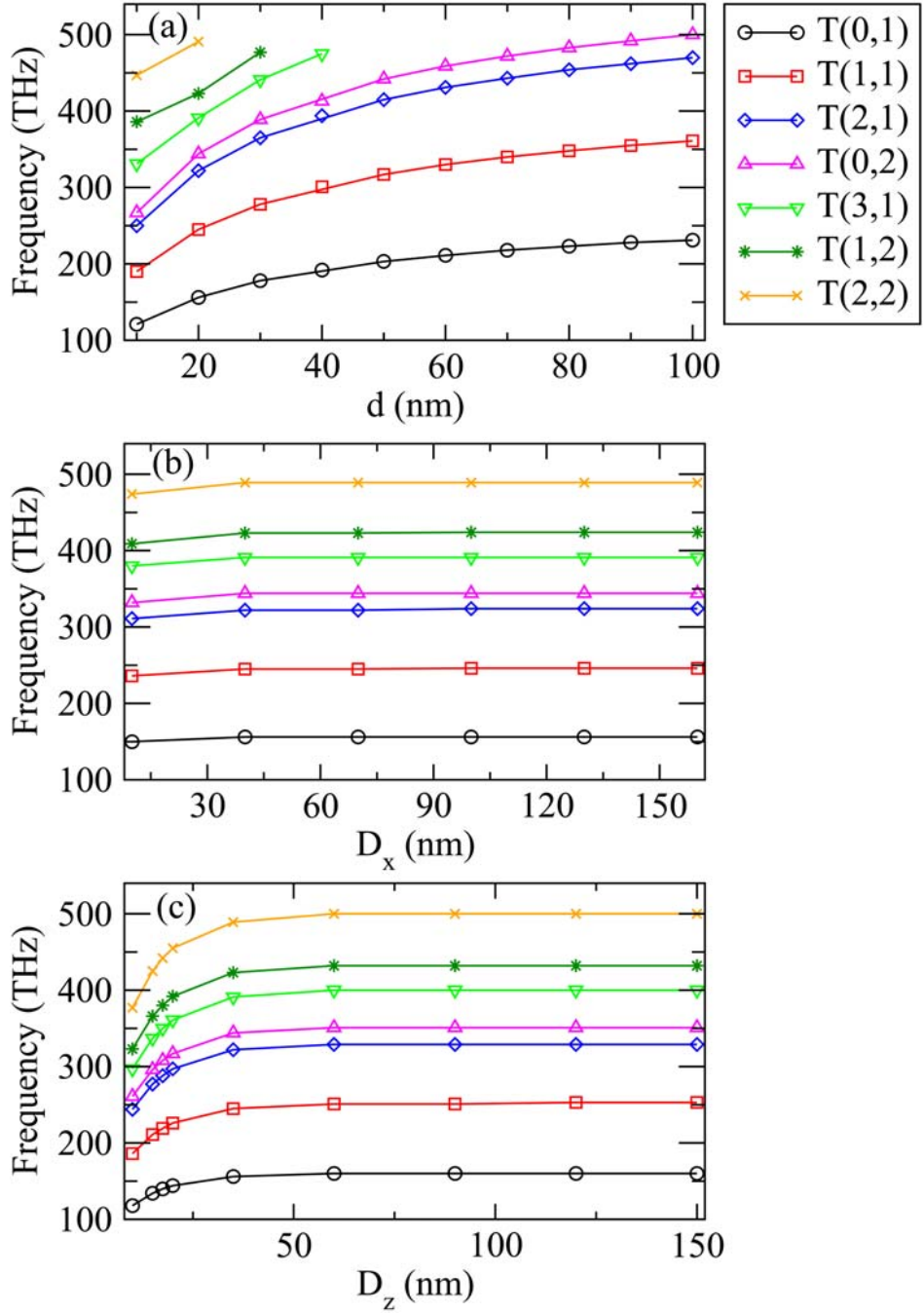

**Figure 6.** The toroid-like mode resonance frequency versus (a) the thickness of the SiO$_2$ layer, (b) the thickness of the sidewall metallic coating, and (c) the thickness of the metallic layer in the $z$ direction. The other parameters are the same as in Figure 2(a), with only the indicated parameter varying in each panel.

# Supporting Information for "Core-Shell Structured Dielectric-Metal Circular Nanodisk Antenna: Gap Plasmon Assisted Magnetic Toroid-like Cavity Modes"

Qiang Zhang[1], Jun Jun Xiao[1,]*, Xiao Ming Zhang[1], Dezhuan Han[2], and Lei Gao[3]

[1]College of Electronic and Information Engineering, Shenzhen Graduate School, Harbin Institute of Technology, Shenzhen 518055, China

[2]Department of Applied Physics, Chongqing University, Chongqing 400044, China

[3]Jiangsu Key Laboratory of Thin Films, School of Physical Science and Technology, Soochow University, Suzhou 215006, China

*Corresponding author: eiexiao@hitsz.edu.cn

**This file includes**:

**SECTION 1**: **Results for the other two excitation configurations**

**SECTION 2**: **Current density distribution in CSND and CMDMR**

**SECTION 3**: **Results for PEC sidewall coating**

**SECTION 4**: **Fano effect from scattering interference between the exterior SPP resonance and the gap SPP resonance**

**SECTION 5**: **ACS versus the radius of the dielectric disk**

**SECTION 6**: **Absorption spectra for different thickness of the metal layer**

**SECTION 7**: **Dipole source decay rate modified by the open and close cavities**

**References**

## Section 1: Results for the other two excitation configurations

For incident plane wave polarized along the $y$-axis in an end-fire configuration (Figure S1(a)), the optical spectra are shown in Figure S1(b). In this situation, the scattering cross section (SCS) dominates the extinction cross section (ECS) and three peaks appear at $f = 190$ THz, 319 THz and 443 THz in the SCS. From the electric field $E_y$ shown in Figure S1(c), it is clearly seen that these modes are correspondingly electric dipole, electric quadrupole, and electric hexapole. The three peaks in Figure S1(b) come from the contributions of these resonant electric multipoles which can be excited by an end-fire plan wave due to the retardation effect, just as in the case of a single silver nanodisk.

For normal incident plane wave as shown in Figure S2(a), the optical cross sections are dominated by a broad resonance whose peak frequency is $f \approx 199$ THz. However, in the absorption cross section (ACS) shown in the insert of Figure S2(b) there are two sharp peaks visible at $f = 245$ THz and 423 THz in the shoulder of this broad resonance spectrum. The $E_z$ distribution and the magnetic field pattern for these three peaks are shown in Figures S2(c) and S2(d), respectively. For the first mode at $f = 199$ THz, $E_z \approx 0$ in the middle dielectric layer. This peak does not correspond to any cavity SPP mode but the electric plasmonic dipole mode that also appear as the first peak (~190 THz) in Figure S1(b). The slight frequency different is due to the retardation effect. For the other two resonances, the near field shown in Figures S2(c) and S2(d) indicate that they are respectively toroid-like modes **T**(1,1) and **T**(1,2).

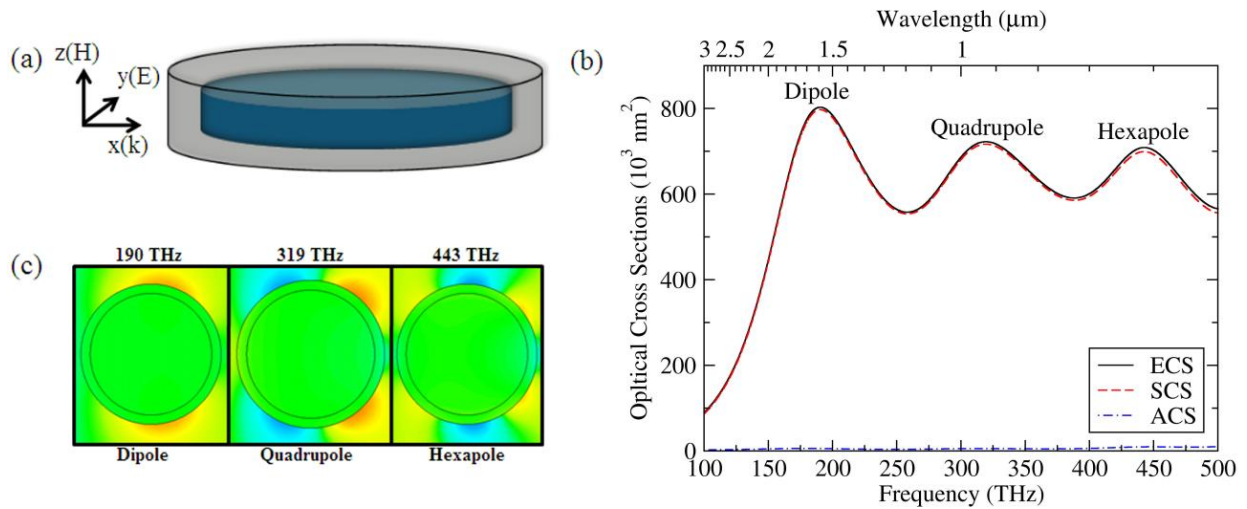

Figure S1. (a) The end-fire configuration. (b) Optical cross sections. (c) $E_y$ pattern at resonances.

The necessary condition to excite the cavity SPP modes is that the incident wave has its magnetic field component across the gap, i.e., one must ensure nonzero incoming magnetic field in the $xy$ plane. In addition, when the plane wave is normally incident to excite the SPP cavity modes, the **T**-mode node can change only in the polarization direction. For oblique incidence of both s- and p-polarizations, the excited resonance is simply combination of the three orthogonal cases we have shown.

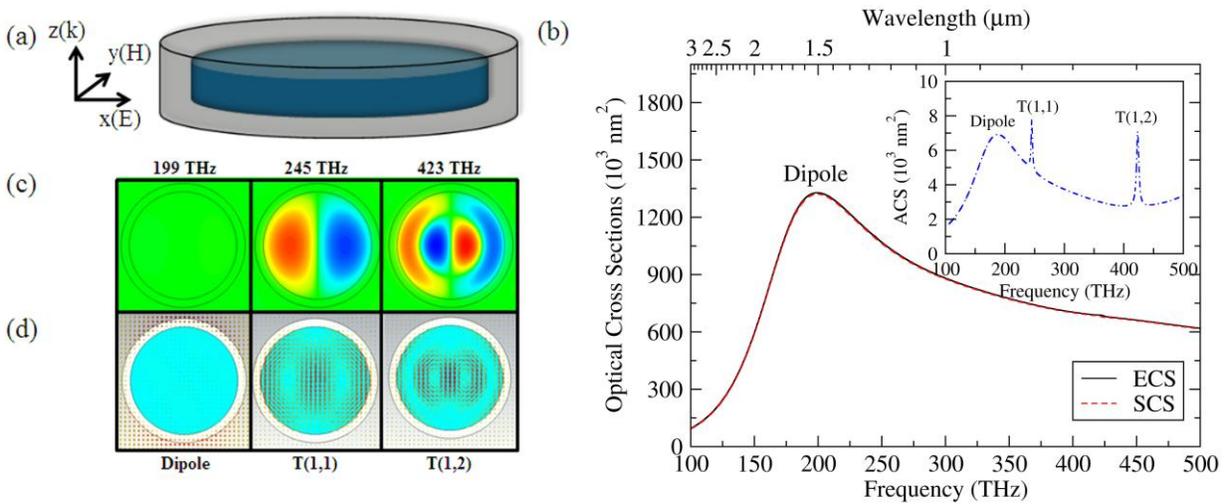

Figure S2. (a) Normal incidence configuration. (b) Extinction and scattering cross sections. Inset shows the absorption cross section with visible gap modes. (c) $E_z$ and (d) magnetic field at resonances.

## Section 2: Current density distribution in CSND and CMDMR

To see the difference of the induced current in the CSND and the CMDMR, we plot the current density of the mode **T**(2,1) in CSND and the mode M(2,1) in CMDMR in Figures S3(a)-(b) (at cross plane $z = 27.5$ nm) and in Figure S3(c)-(d) (at cross plane $y = 0$ nm). For the mode **T**(2,1) in CSND, the induced currents oscillate in an actinomorphic pattern like a breathing mode. This kind of current distribution is very similar to the current flowing on the surface of a torus. Another important feature is that the ends (zero value) of the in-plane currents are at specific interior points of the silver disks [see Figure S3(a)]. While for the mode **M**(2,1) of the CMDMR, the current always starts from one edge point and terminate at another edge point [see Figure S3(b)]. The presence of sidewall metallic coating is therefore to prohibit the currents being terminated at the edge. On the contrary, it forces the conduction current flowing to the opposite disk along the sidewall coating layer [see Figure S3(c)]. Due to these apparent differences of the current distribution, CSND can only sustain toroid-like cavity modes while the CMDMR support additional magnetic multipole cavity modes.

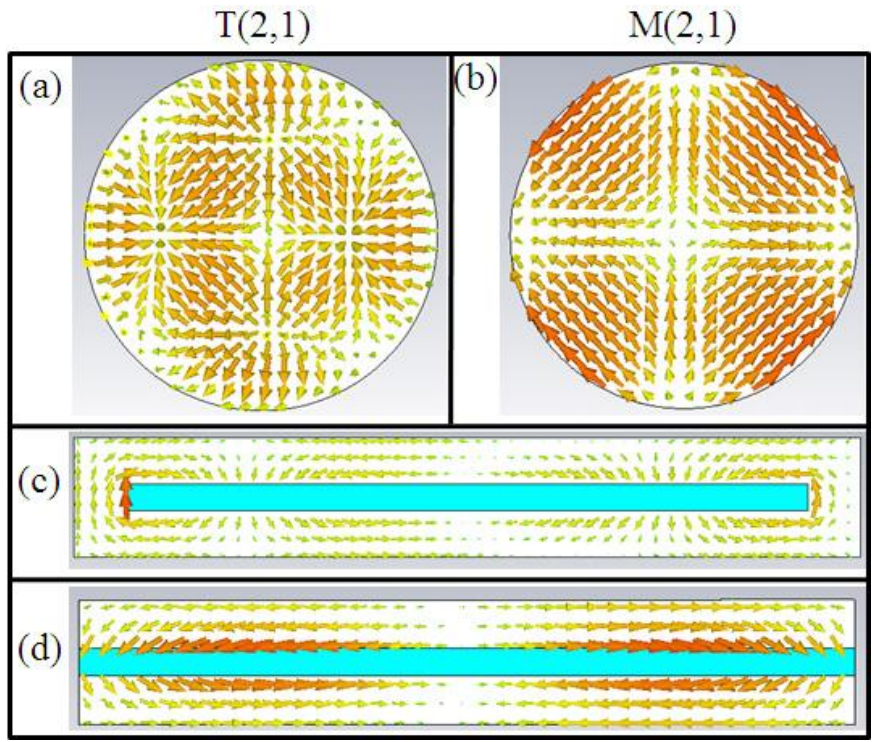

Figure S3. Current density vector of (a) the mode **T**(2,1) in CSND and (b) the mode M(2,1) in CMDMR across *xy* plane in the silver layer at $z = 27.5$ nm. (c) and (d) are in the *xz* plane for $y = 0$ nm.

## Section 3: Results for PEC sidewall coating

Figure 4S shows that the cavity mode peaks also appear in the ACS spectrum for the case of PEC sidewall coating. By examining the corresponding near field distributions (figures not shown here), it is verified that these resonances are the toroid-like cavity modes, the same as in the case of metallic coating.

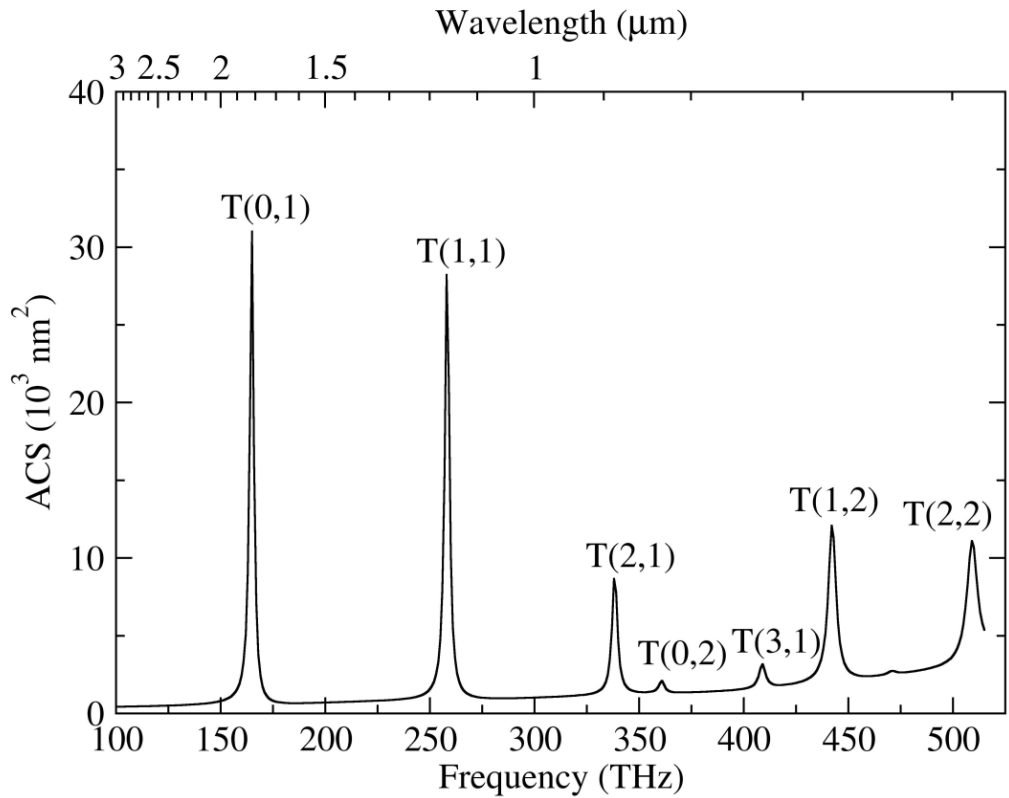

Figure S4. ACS spectrum for CSND with PEC sidewall coating.

Figure S5 shows the $E_z$ field for **T**(0,1) and **T**(1,1) along a radial direction. Here PEC coating allows exactly vanishing $E_z$ in the coating domain (the regime between the vertical blue and green dashed lines). In this situation, the boundary condition at the edges of the dielectric layer is strictly $E_z(\rho = R) = 0$.

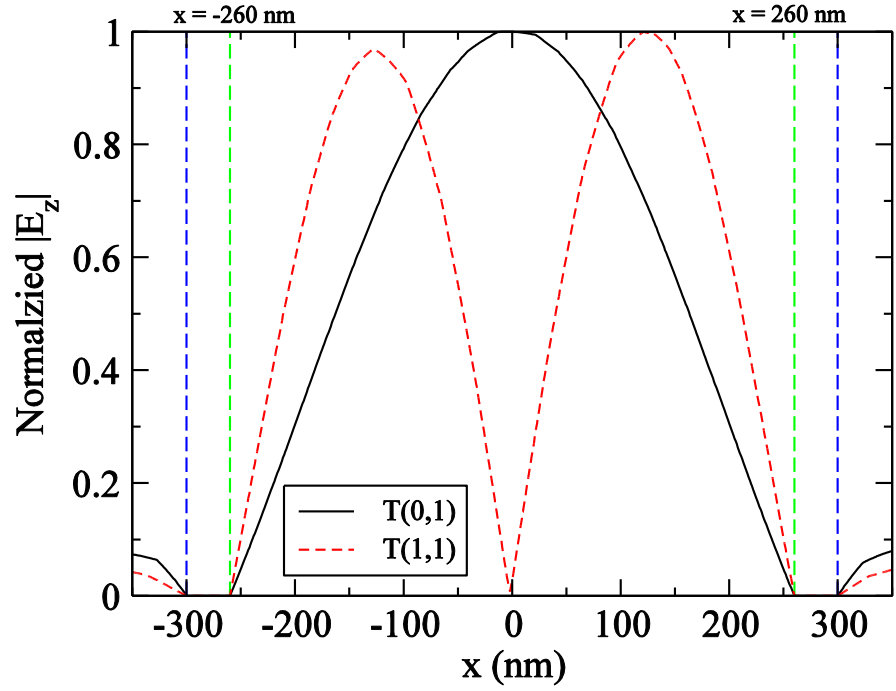

Figure S5. $E_z$ field (magnitude) along the *x* direction in the *xz* plane for mode **T**(0,1) and **T**(1,2). The dashed vertical green lines mark the boundary of the dielectric layer while the dashed vertical blue lines define the boundaries of the individual CSND.

**Section 4: Fano effect from scattering interference between the exterior SPP resonance and the gap SPP resonance**

Figure S6(a) plots the ECS spectra of a solid silver disk without the dielectric core (keep the geometry size), of the CSND, and of the CMDMR in the frequency window of 500 THz-1000 THz. It is clearly seen that the three structures have dominated resonance at $f = 733$ THz, 739 THz and 762 THz, respectively. The second dominated resonance appears near $f = 830$ THz in all the ECS curves. On the other hand, there are several sub-peaks on the left shoulder of the main peak for both CSND and CMDMR but not for the solid disk (black line). To identify these resonances we present the corresponding $H_y$ and $E_z$ field in Figures S6(b) and 6(c), respectively. Firstly we focus on the resonance at $f = 733$ THz, from the $H_y$ field shown in Figure S6(b) we see typical SPP wave propagating at the outer surfaces (exterior SPP) of the metallic part. This exterior SPP exist in all the three structures (i.e., solid disk, CMDMR, and CSND). With the dielectric core in the CSND and the CMDMR, a gap SPP wave can be excited simultaneously in the core layer, although it is much weaker than the exterior SPP. Particularly, we have chosen the resonant gap SPP at $f = 545$ THz in the CSND and that at $f = 557$ THz in the CMDMR. The gap SPP resonant mode has much stronger $H_y$ field in the gap region than that of the exterior SPP mode, as seen in Figure S6(b). Moreover, the strong $E_z$ field shown in Figure S6(c) can help to identify them as the gap cavity modes **T**(3,2) and **T**(2,3) respectively. Notice that on the gap SPP resonances, the exterior SPP wave is still there but far-off the exterior SPP main resonance. Because the gap cavity modes have relatively higher quality factor comparing with the exterior SPP mode, the destructive interference can happen between their scatterings, resulting in asymmetric Fano line shape near the gap SPP resonances. The second dominated resonance near $f = 830$ THz comes from the higher order exterior SPP mode. Notice that more field nodes (visibly 5) than the first exterior SPP resonance (around 4 nodes) appear in the $H_y$ pattern of the pure solid disk, as shown in Figure S6(b). The node number in the outer surface can be more clearly seen in the current density pattern. The exterior SPP can also be clarified by the current density at their resonance frequencies. Figure S7 shows that the currents tend to concentrate on the interfaces between the metal shell and the background.

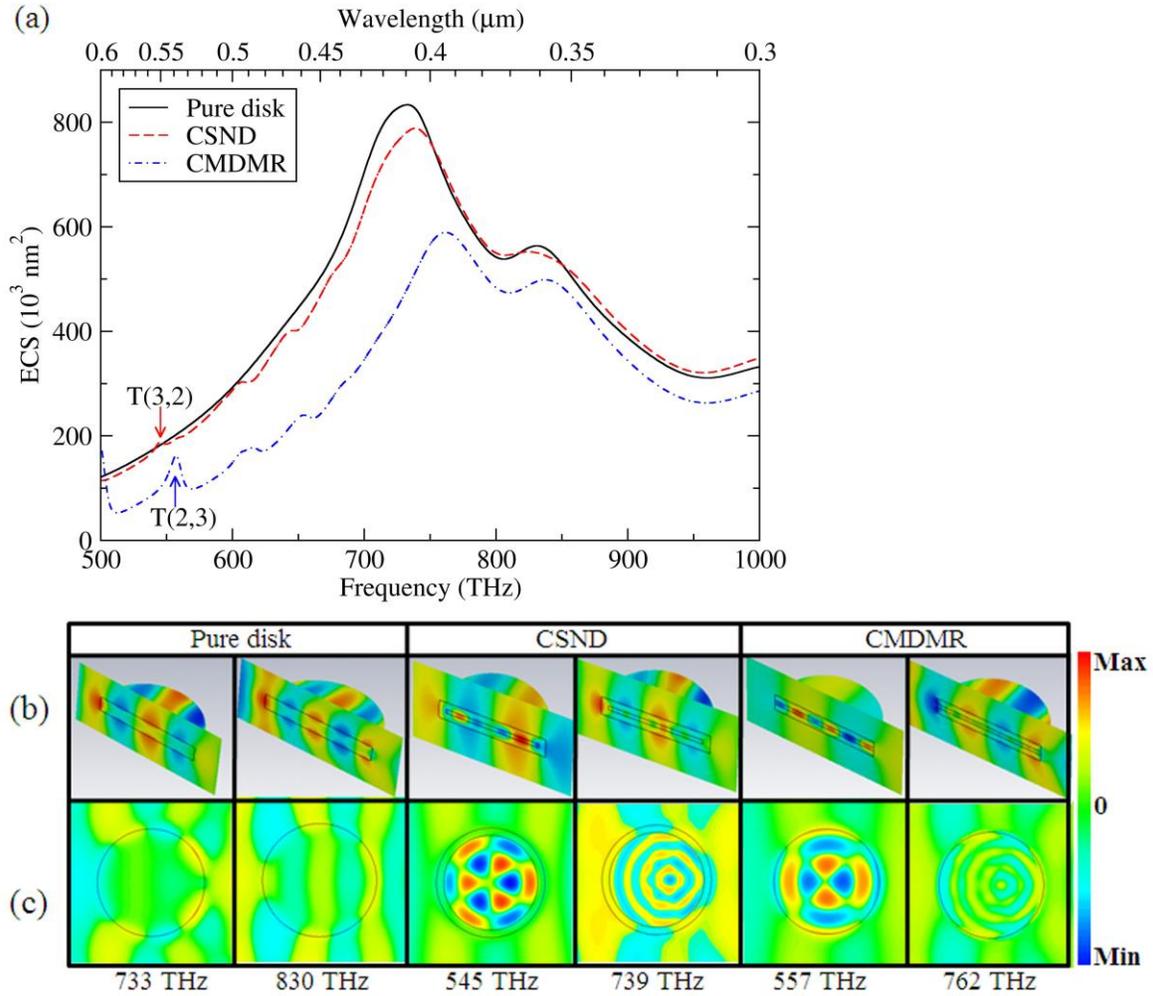

Figure S6. (a) ECS spectra for three different kinds of structures (e.g., solid disk, CMDMR, and CSND). The red arrow indicates the **T**(3,2) mode in the CSND at $f = 545$ THz. The blue arrow indicates the **T**(2,3) mode in the CMDMR at $f = 557$ THz. (b) $H_y$ field for the three structures at their typical resonances. (c) $E_z$ field in the middle slice ($z = 0$) of the corresponding resonances.

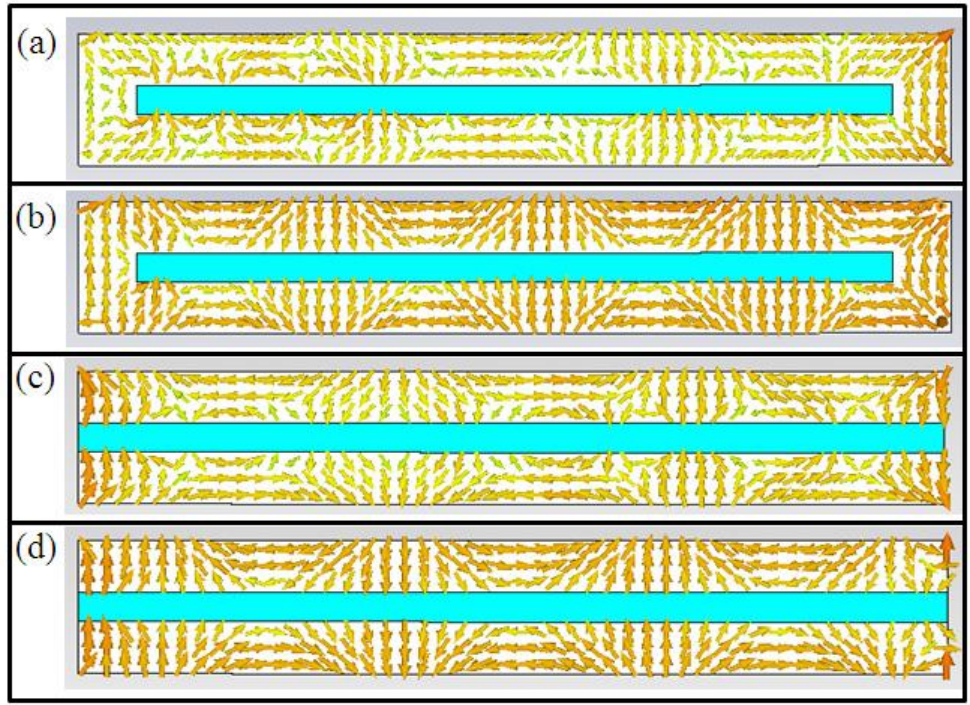

Figure S7. Current density in the $xz$ plane ($y=0$) of the CSND at (a) $f=738$ THz; (b) $f=830$ THz, also of the CMDMR at (c) $f=766$ THz; (d) $f=830$ THz.

## Section 5: ACS versus the radius of the dielectric disk

Figure S8 shows the absorption spectrum of the CSND with various $R$, in the considered frequency band from 100 THz to 500 THz. As the size of the dielectric disk increases, more absorption bands emerge, originated from the higher-order cavity modes. The lower-order modes are well discernible in Figure S8. For example, from low frequency to high frequency the first two absorption bands are of **T**(0,1) and **T**(1,1). However, certain higher-order modes in the high frequency regime become difficult to be distinguished due to the weak excitation.

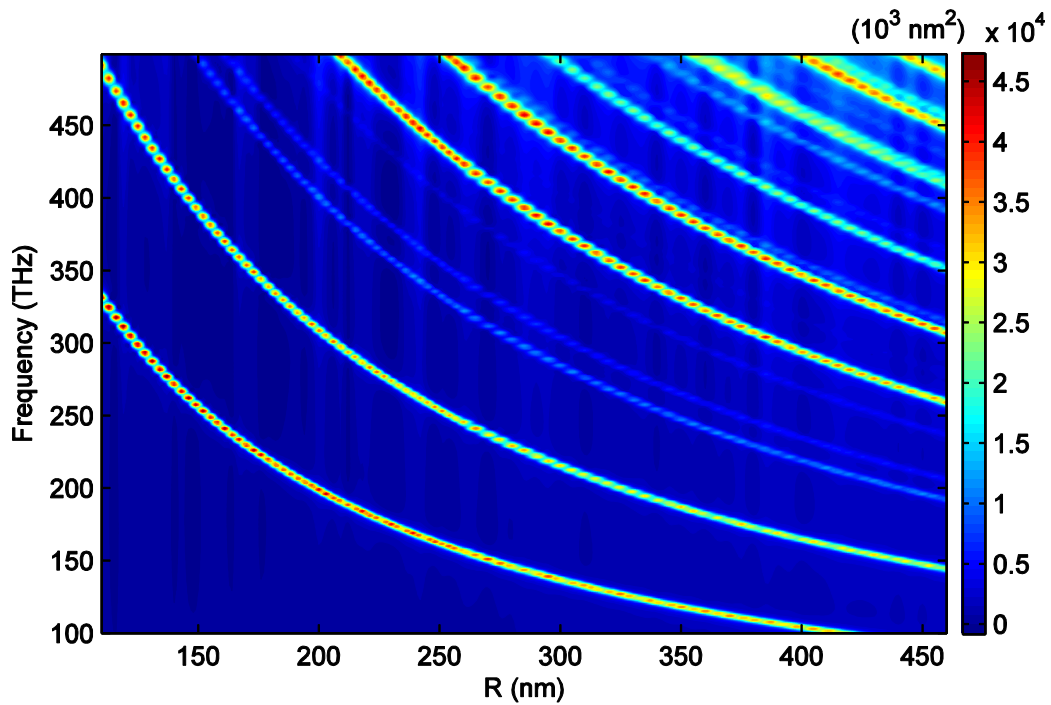

Figure S8. ACS for varying radius of the dielectric disk in the frequency band from 100 THz to 500 THz.

## Section 6: Absorption spectra for different thickness of the metal layer

As a matter of fact, the dispersion relation described by Eq. (2) is valid only when the thickness of the metal layer $D_z$ is large enough. However, as $D_z$ increases, some of the toroid-like cavity modes become difficult to couple with the external plane wave. This can be seen in Figure S9 which shows that when $D_z = 60$ nm, the ACS peak of **T**(3,1) mode, which originally locates at $f = 400$ THz ((blue dash-dotted line, pointed marked by the downward arrow), become too weak to be observed.

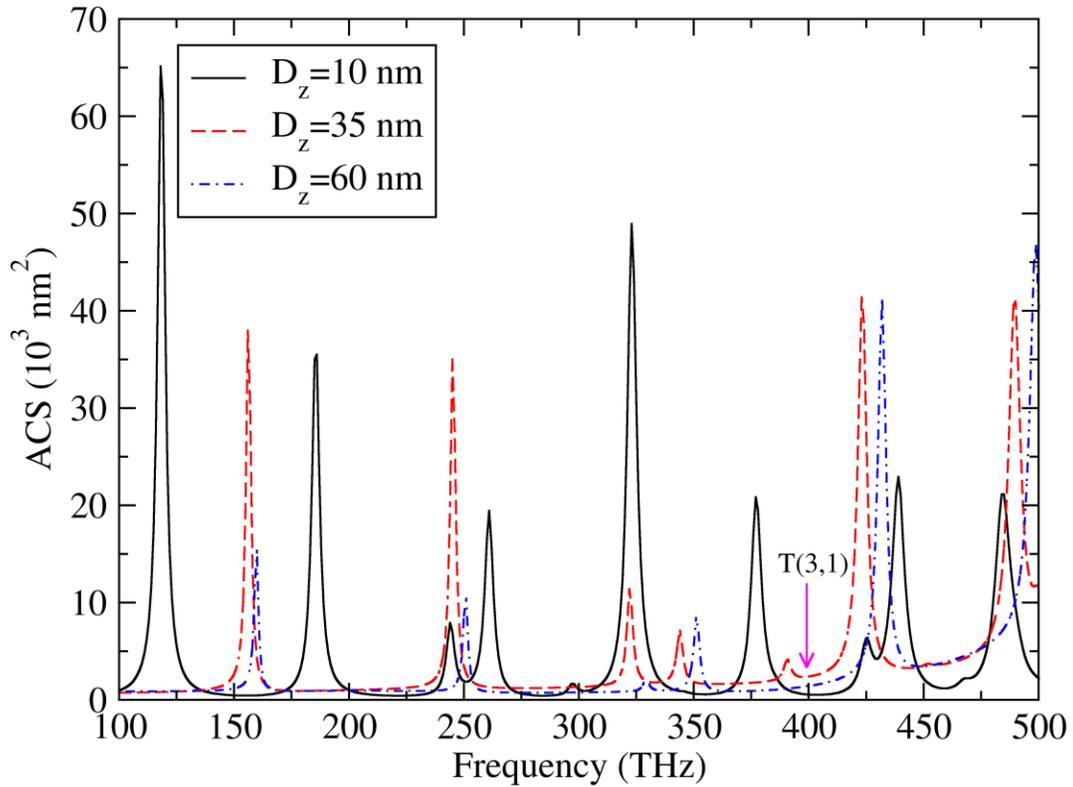

Figure S9. The ACS spectra for different thickness of the metal layer in $z$ direction. The downward arrow marks the position of the **T**(3,1) mode for $D_z = 60$ nm.

## Section 7: Dipole source decay rate modified by the open and close cavities

To numerically explore the effects of these SPP-associated modes on the photoluminescence (PL) of emitting material (for example, MEH-PPV) coated nearby, we use an electric point dipole to model the emission of the luminescent molecule and assume the weak-coupling regime [1-4]. The configuration is shown in Figure S10. We calculate the radiative decay rate ($\gamma_r$) and the nonradiative decay rate ($\gamma_{nr}$) of the dipole at different positions P$_1$ and P$_2$ near the CSND and the CMDMR structures. The normalized decay rate $\gamma_r$ and $\gamma_{nr}$ are given by $\gamma_r = P_r / P_0$ and $\gamma_r = P_{nr} / P_0$, where the radiative power $P_r$, the nonratiative power $P_{nr}$ and the radiative power of the dipole are obtained via [2]

$$P_r = \frac{1}{2} \text{Re}[\int_{S_r} (\mathbf{E} \times \mathbf{H}^*) \cdot d\mathbf{s}],$$

$$P_{nr} = -\frac{1}{2} \text{Re}[\int_{S_{nr}} (\mathbf{E} \times \mathbf{H}^*) \cdot d\mathbf{s}],$$

here $\varepsilon_0$, $c$ and $\mathbf{p}_0$ are respectively the vacuum permittivity, light speed in vacuum and the dipole

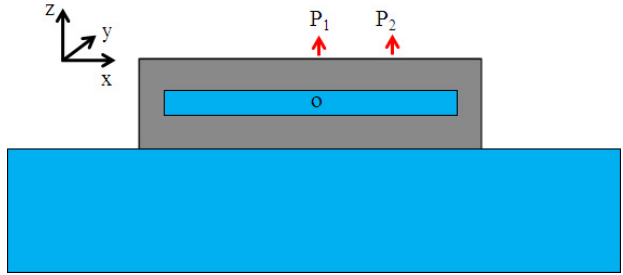

Figure S10. Schematic illustration of the CSND interacting with an electric point dipole. The geometry of the CSND is the same as in Fig. 2. A substrate is appended here whose size is set as $1 \times 1 \times 0.15 \ \mu m$. The dipole emitter is 5 nm away from the top surface of the CSND and polarized vertically (red arrows). We consider two different positions P$_1$ and P$_2$ whose $x$-coordinate is 0 nm and 130 nm, respectively. The same configuration applies to the CMDMR structure.

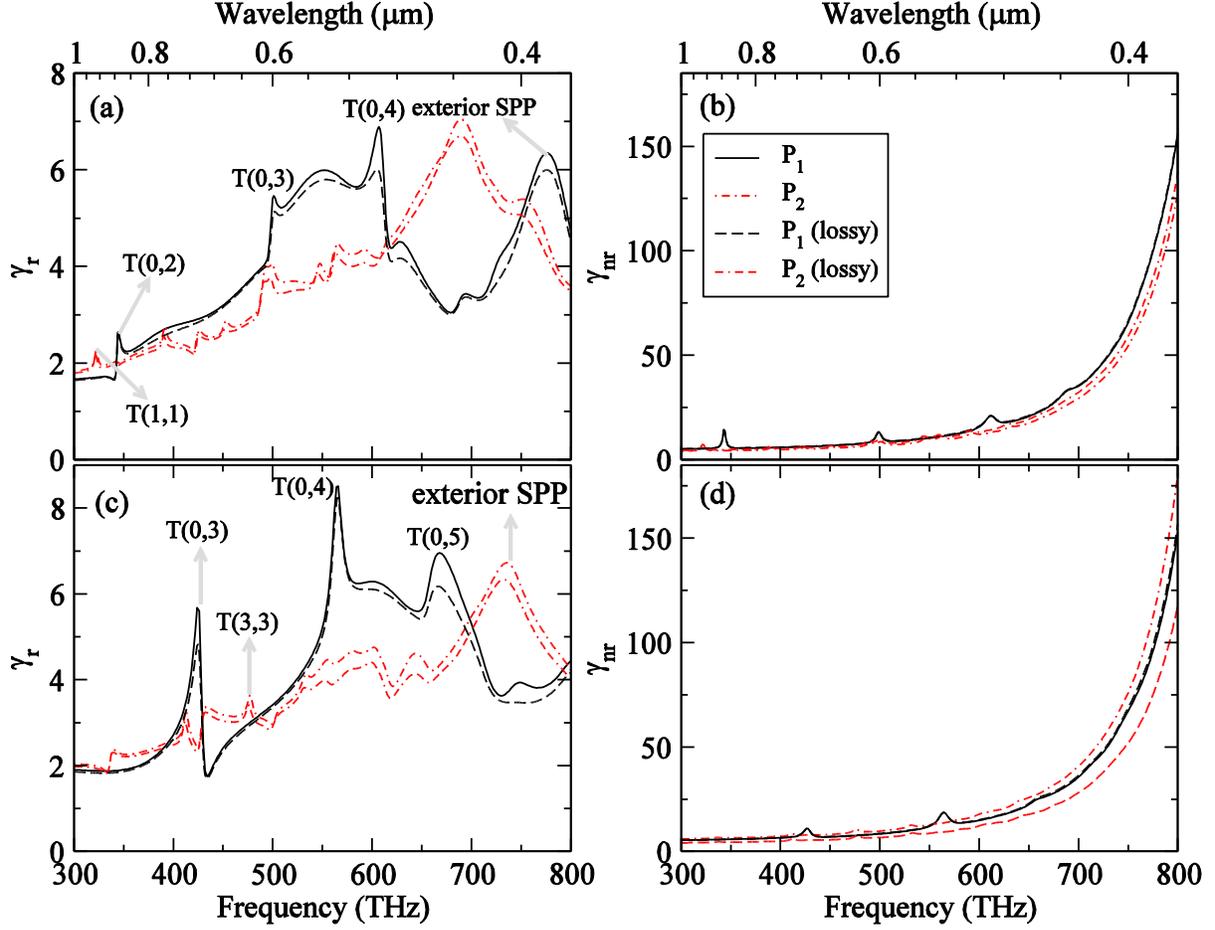

Figure S11. (a) $\gamma_r$ and (b) $\gamma_{nr}$ in the CSND (close cavity). (c) $\gamma_r$ and (d) $\gamma_{nr}$ in CMDMR (open cavity). Black curves are for $P_1$ and red lines are for $P_2$.

moment. The vacuum decay rate is $P_0 = \omega^4 |\mathbf{p}_0|^2 / 12\pi\varepsilon_0 c^3$.

The results for CSND and CMDMR are shown in Figures S11(a)-S11(b) and Figures S1s(c)-S11(d), respectively. Firstly, it is seen that $\gamma_{nr}$ is overall much greater than $\gamma_r$ and the resonance peaks in the $\gamma_{nr}$ is hard to be distinguished (but still can be seen). On the contrary, the peaks are clearly seen in the $\gamma_r$ spectrum which indicates that the SPP resonances of the structures enhance the dipole emission in both the CSND or in CMDMR. Furthermore, the obvious differences between the line shapes of $P_1$ and $P_2$ demonstrate that the position of the dipole emitter dramatically affects its radiation characteristics. By carefully examining the near field distributions, for the gap modes it is found that only the modes with zero azimuthal number can be excited when the dipole source is at the center axis, e.g., at $P_1$. As the dipole is moved to $P_2$, all the gap SPP resonances can be excited but their relative strength could be quite different. The degree of field overlap and symmetry matching between the source and the mode is crucial for the excitation efficiency. This explains that more resonance peaks are seen in the curves for $P_2$ (red lines in Figures S11), as compared to those of $P_1$. Here we focus on some typical mode from the $\gamma_r$ spectra in Figure S11(a) and Figure S11(c) (marked by the gray arrows with their corresponding mode names). It is noticed that the main peaks in $\gamma_r$ actually come from the extior SPP mode because they radiate more easily. In addtiation, the resonance frequency of the gap SPP mode is insensitive to the substrate and the positions of the dipole emitter while those of the exterior SPP modes are obviously affected by the substrate and the position of the dipole. The electric fields of several resonance excitations are shown in Figure S12. It is seen that the gap SPP modes have apparent interference patterns which are not visible in the exterior SPP modes [Figures S12(c) and S12(f)]. As shown in the $xz$ plane, when the frequency approaches to $f \approx 750$ THz, typical SPP waves are generated at the interfaces between the metal shell and the air, as well as the metal shell and the substrate.

Finally, in order to evaluate the effect of absorption in the substrate on relevant luminescent spectrum, we deliberately add certain loss to the substrate by setting the imaginary part of its permittivity to 0.1. As shown in Figure S11 (dash-dotted lines), no remarkable influence on the decay rate is observed, except for the slight strength modification on $\gamma_r$ and $\gamma_{nr}$.

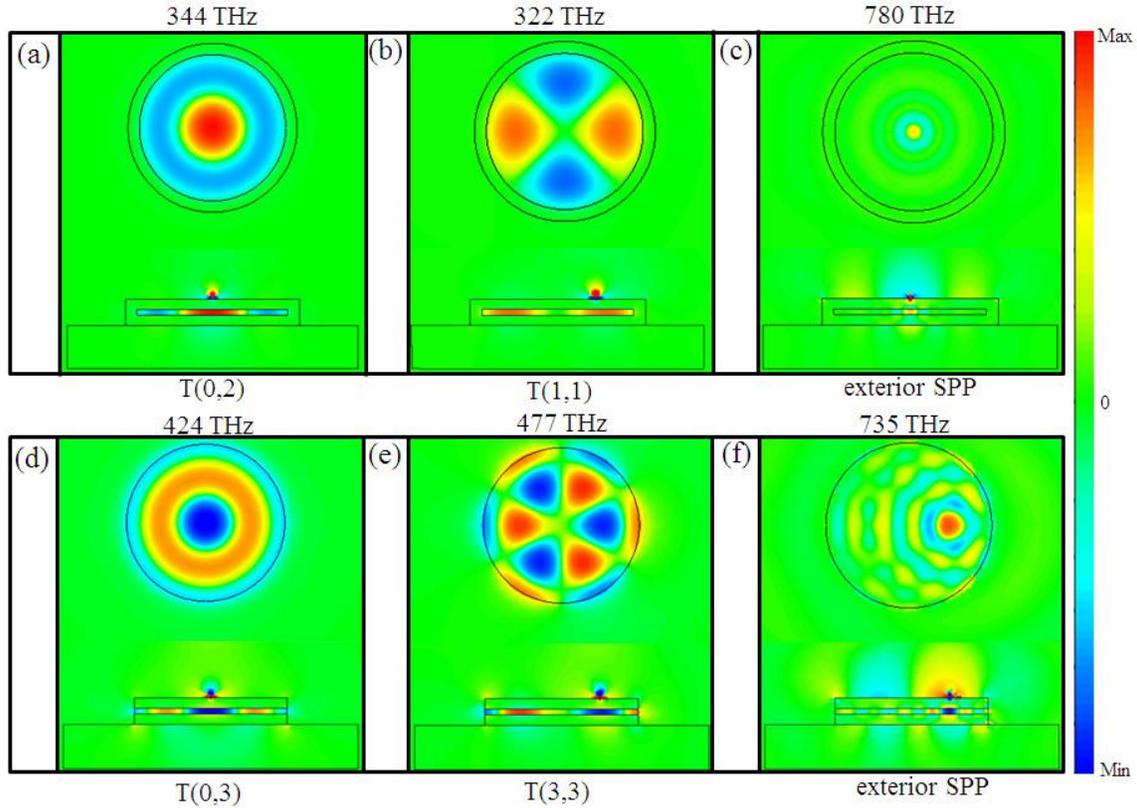

Figure S12. Near field ($E_z$) of chosen modes under dipole source excitation in presence of a substrate. Inside each panel the upper is for $xy$ plane, and the bottom for $xz$ plane.

**References**:

[1] Salomon, A.; Gordon, R. J.; Prior Y.; Seideman, T.; Sukharev, M. Strong coupling between molecular excited states and surface plasmon modes of a slit array in thin metal film. *Phys. Rev. Lett.* **2012**, *109*, 073002.

[2] Zhang, X. M.; Xiao, J. J.; Zhang, Q. Interaction between single nano-emitter and plasmonic disk-ring nanostructure with multipole Fano resonances. *J. Opt. Soc. Am. B* **2004**, *31*, 2193-2200.

[3] Russell, K. J.; Liu, T. L.; Cui, S.; Hu, E. L. Large spontaneous emission enhancement in plasmonic nanocavities. *Nat. Photonics* **2012**, *6*, 459-462.

[4] O'Carroll, D. M.; Fakonas, J. S.; Callahan D. M.; Schierhorn, M.; Atwater, H. A. Metal-polymer-metal split-dipole nanoantennas. Adv. Mater. **2012**, *24*, OP136-OP142.